%% file: sample-manuscript.tex
\documentclass[sigconf]{acmart}
\usepackage{array}
\usepackage{enumitem}

\AtBeginDocument{%
  }

\copyrightyear{2026}
\acmYear{2026}
\setcopyright{cc}
\setcctype{by}
\acmConference[COMPASS '26]{ACM SIGCAS/SIGCHI Conference on Computing and Sustainable Societies}{July 27--31, 2026}{Virtual Event, USA}
\acmBooktitle{ACM SIGCAS/SIGCHI Conference on Computing and Sustainable Societies (COMPASS '26), July 27--31, 2026, Virtual Event, USA}
\acmDOI{10.1145/3811242.3819106}
\acmISBN{979-8-4007-2702-3/2026/07}

\begin{document}

\title[Asymmetries in Institutional Accountability and Collective Sensemaking]{``Is This Not Enough?": Asymmetries in Institutional Accountability and Collective Sensemaking in the Case of Canada's Algorithmic Visa Triage System}

\author{Dipto Das}
\affiliation{
  \department{Department of Computer Science}
  \institution{University of Toronto}
  \city{Toronto}
  \state{Ontario}
  \country{Canada}
}
\email{dipto.das@utoronto.ca}
\author{Matthew Tamura}
\affiliation{
  \department{Faculty of Information}
  \institution{University of Toronto}
  \city{Toronto}
  \state{Ontario}
  \country{Canada}
}
\email{matthew.tamura@mail.utoronto.ca}
\author{Syed Ishtiaque Ahmed}
\affiliation{
  \department{Department of Computer Science}
  \institution{University of Toronto}
  \city{Toronto}
  \state{Ontario}
  \country{Canada}
}
\email{ishtiaque@cs.toronto.edu}
\author{Shion Guha}
\affiliation{
  \department{Faculty of Information}
  \institution{University of Toronto}
  \city{Toronto}
  \state{Ontario}
  \country{Canada}
}
\email{shion.guha@utoronto.ca}

\begin{abstract}
  This paper examines how algorithmic accountability in Canada's visa system is articulated institutionally and experienced by applicants across borders. We analyzed Immigration, Refugees and Citizenship Canada (IRCC)'s Algorithmic Impact Assessment (AIA) for the temporary resident visa (TRV) triage system using the algorithmic decision-making adapted for the public sector (ADMAPS) framework and analyzed Reddit discussions among applicants using a mixed-methods approach. We show that while institutional artifacts emphasize transparency, procedural safeguards, and bounded impacts, applicants engage in collective sensemaking to interpret opaque decisions, often relying on peer knowledge amid uncertainty. We identify three asymmetries between how institutional accountability is structured and how people perceive the process: epistemic asymmetry in access to decision logic, jurisdictional asymmetry in exposure shaped by geopolitical positioning, and temporal--relational asymmetry in how waiting and uncertainty are experienced. We emphasize why it is important to shift attention from institutional design to the uneven distribution of experiences with public-sector algorithmic governance. Together, these contributions demonstrate how algorithmic governance systems in the context of transnational migration produce structured asymmetries not captured by institutional disclosure frameworks, and how extending ADMAPS can account for those uneven translations of accountability.
\end{abstract}



\maketitle
\input{sections/introduction}
\input{sections/literature_review}
\input{sections/methods}
\input{sections/results}
\input{sections/discussion}
\input{sections/conclusion}

\begin{acks}
    This study was partially supported by the Institute of Health Emergencies and Pandemics Postdoctoral Fellowship, the School of Cities Urban Challenge Grant at the University of Toronto, and an NSERC Discovery Grant.
\end{acks}

\bibliographystyle{ACM-Reference-Format}
\bibliography{sample, sample-base}

\end{document}

%% file: sections/introduction.tex
\section{Introduction}
Immigration, Refugees and Citizenship Canada has been using a tool known as Chinook since 2018 to rapidly process visa applications by summarizing complex files into simplified views, allowing decisions to be made much more quickly~\cite{steinman2023chinook}. Immigration lawyers and advocates raised concerns that this efficiency could come at the cost of fairness, as officers sometimes relied on condensed information rather than full application reviews, citing the use of standardized refusal templates and generic reasoning, which made decisions appear ``very generic" and disconnected from applicants' submitted evidence~\cite{mach2026lack}. IRCC maintains that Chinook is not an AI system but rather a Microsoft Excel-based tool that does not make decisions itself~\cite{gc2022cimm}. However, the lack of transparency in its decision-making processes has fueled ongoing debate about accountability in immigration processing~\cite{chartier2025ai}. IRCC has also deployed the Advanced Analytics Triage of Overseas Temporary Resident Visa (TRV) Applications, a distinct system that algorithmically triages applications by risk level to prioritize processing, thereby shaping how applications enter and are subsequently handled. Public controversies surrounding immigration technologies, such as the one regarding Chinook, highlight persistent tensions between institutional claims of transparency and applicants' ability to understand and navigate these systems. As public institutions increasingly deploy algorithmically supported decision-making~\cite{saxena2024algorithmic, moon2025datafication, redden2022automating}, the computing scholarship on sustainable societies should examine social perception around algorithmic governance systems.

In Canada, the Directive on Automated Decision-Making (ADM)~\cite{gc2021adm} and the Algorithmic Impact Assessment (AIA)~\cite{gc2026aiatool} are intended to make such systems legible and accountable to external audiences. While governance frameworks proposed and practiced by such entities promise transparency, accountability, and procedural fairness~\cite{NRC_visaalgo_review}, these are largely driven by institutional, technical, or policy-level perspectives, including audits, documentation frameworks, and impact assessments~\cite{goodman2022algorithmic, metcalf2021algorithmic, watkins2021governing}. These approaches conceptualize accountability in terms of traceability, disclosure, and procedural safeguards, often assuming that increased transparency leads to improved understanding and oversight~\cite{kroll2021outlining}. However, there is limited empirical work examining how institutional accountability artifacts relate to the lived experiences of individuals who are subject to and navigate these systems. This gap becomes particularly pronounced in transnational domains, such as immigration, where institutional logics are uneven, and decisions carry significant consequences. To address this gap, we adopt a comparative perspective that reads institutional accountability alongside applicant experience. 

If accountability is articulated through structured disclosure about algorithms but experienced through partial visibility around evaluative criteria, then interpretability must be reconstructed in practice~\cite{ananny2018seeing, wachter2017counterfactual}. In such contexts, collective sensemaking emerges as a necessary response to algorithmic opacity~\cite{weick2005organizing, mohlmannn2023algorithm}. We examine this through a case study of IRCC's automated triage system for TRV applications. We analyze two complementary data sources: (1) IRCC's Algorithmic Impact Assessment (AIA) and a National Research Council (NRC) peer review, which together articulate institutional accountability and system design, and (2) a corpus of Reddit discussions, which capture applicants' accounts of navigating the visa process. We map institutional artifacts onto the algorithmic decision-making adapted for the public sector (ADMAPS) framework to analyze how accountability is structured across bureaucratic processes, human discretion, and algorithmic decision-making, and we conduct a mixed-methods analysis of Reddit discussions to examine how applicants interpret and respond to system outcomes. This enables us to identify points of alignment and divergence between institutional articulation and experiential interpretation.

Our findings reveal three interrelated asymmetries. First, an \emph{epistemic asymmetry}, in which institutional claims of transparency and interpretability do not translate into accessible understanding for applicants, who instead reconstruct decision logic through peer exchange. Second, a \emph{jurisdictional asymmetry}, in which applicants' experiences are shaped by their positioning within transnational mobility regimes, despite institutional framing of uniform procedural governance. Third, a \emph{temporal--relational asymmetry}, in which waiting and delay function as consequential forms of exposure that are not adequately captured in institutional impact assessments. Taken together, these findings highlight a limitation in existing frameworks around ADM in the public sector and motivate an extension of ADMAPS to account for transnational contexts.

%% file: sections/literature_review.tex
\section{Literature Review}
This section examines algorithmic governance as an administrative and sociotechnical phenomenon, tracing how institutional mechanisms for transparency and accountability are structured, limited in practice, and subsequently reinterpreted through collective sensemaking. We then situate these dynamics within immigration and visa processing--contexts that are inherently transnational.

\subsection{Public-Sector Algorithmic Governance as Administrative Experience}
Algorithmic decision-making systems have become increasingly embedded within public-sector governance, where they are used to allocate resources, assess eligibility, detect risk, and triage applications across domains such as social services, policing, healthcare, and immigration~\cite{eubanks_automating_2018, zajko_automated_2023, angwin2016machine}. In these contexts, algorithmic systems do not operate as standalone technical tools but as components of broader administrative infrastructures, shaping how decisions are made, how cases are processed, and how individuals are classified and acted upon~\cite{redden2018democratic}. As such, public-sector algorithmic systems are best understood as sociotechnical arrangements that redistribute decision-making across bureaucratic procedures, human discretion, and computational processes~\cite{saxena2023rethinking}.

A growing body of work has examined how these systems transform governance by reconfiguring authority, responsibility, and accountability within administrative institutions~\cite{alon2024public}. Rather than eliminating discretion, algorithmic systems often shift it across actors and stages, embedding decision-making within data pipelines, classification schemes, and organizational routines~\cite{busuioc2021accountable, de2020artificial}. These shifts can obscure lines of responsibility and complicate oversight, particularly when decision-making is distributed across multiple actors~\cite{cobbe2023understanding, kuziemski2020ai}. In the public sector, legal mandates, policy instruments, and institutional constraints often seek to formalize fairness, accountability, and transparency through procedural requirements. For instance, Canada's Algorithmic Impact Assessment (AIA) evaluates algorithmic systems through a structured questionnaire composed largely of predefined and often binary or scalar prompts~\cite{gc2026aiatool}, such as whether a system influences access to rights or services, relies on sensitive personal data, produces legally or materially significant outcomes, or incorporates human oversight and recourse mechanisms. Responses are assigned weighted scores and aggregated into impact levels (I–IV), which in turn determine the scope of required governance interventions. Encoding assumptions about risks, these categories shape what aspects of algorithmic systems become visible or actionable within institutional processes.

Foundational work on classification systems and bureaucratic infrastructures demonstrates how seemingly neutral processes organize social life by categorizing individuals, standardizing information, and structuring access to resources and rights~\cite{bowker_sorting_1999, bucher2018if}. Hence, administrative systems are not only sites of institutional decision-making but also lived infrastructures that shape how individuals experience governance in practice~\cite{brodkin2011policy, lipsky2010street}. The recent adoption of algorithmic decision-making in governance produces further experiences of uncertainty and opacity, particularly for populations subject to eligibility assessments, risk scoring, or discretionary review~\cite{eubanks_automating_2018, redden2020datafied, redden2022automating}.

While governance frameworks seek to institutionalize accountability through structured disclosure~\cite{raji2020closing, selbst2019fairness, diakopoulos2016accountability}, less attention has been paid to how these mechanisms are encountered and interpreted by those subject to automated decisions in practice~\cite{wang2020factors, starke2022fairness}. A key dimension of this experience is temporal~\cite{susser2022decision}: administrative processes are often structured through waiting, delay, and asynchronous decision-making, which shape how individuals plan, anticipate, and respond to institutional outcomes. Empirical studies worldwide have highlighted how these temporal dynamics are frequently intertwined with relational consequences, as administrative decisions affect not only individuals but also their families, livelihoods, and mobility trajectories~\cite{sigona2012have, auyero2020patients, hage2009waiting}. In the Canadian context, public-sector adoption of algorithmic systems across domains such as employment services, border management, and immigration further underscores how automated decision-making has become embedded within administrative governance~\cite{gc2025register}. Hence, computing scholarship needs to examine how algorithmic decision-making in such high-stakes administrative contexts, where decisions have significant implications for individuals' rights, status, and life opportunities, is experienced by those subject to them and relate to formal accountability mechanisms.

\subsection{From Institutional Transparency to Collective Sensemaking Under Opacity}
Within policy and technical domains, accountability is often framed as the ability to trace, justify, and assign responsibility for decisions and their outcomes~\cite{binns2018algorithmic, wieringa2020account}, while transparency is understood as the provision of information that enables explanation, documentation, or disclosure of system properties~\cite{balasubramaniam2023transparency}. Governance instruments such as AIAs, documentation frameworks, audit mechanisms, and public reporting infrastructures aim to exercise those principles and render algorithmic systems more legible to regulators, practitioners, and affected populations~\cite{mitchell2019model, sokol2020explainability, felzmann2019transparency}.

However, a growing body of scholarship has critically examined the limits of these approaches, arguing that transparency and accountability are not merely technical properties but sociotechnical practices shaped by institutional logics, organizational constraints, and power relations~\cite{ananny2018seeing, selbst2019fairness, suchman2002located}. From this perspective, disclosure does not straightforwardly translate into understanding. Instead, it is mediated by decisions about what information is made visible, how it is structured, and for whom it is intended. As a result, transparency initiatives often produce forms of legibility that are oriented toward institutional actors, such as auditors, regulators, and technical experts, while remaining inaccessible or insufficient for those directly affected by algorithmic decisions~\cite{eyert2023rethinking, norval2022disclosure}. Documentation frameworks, reporting templates, and assessment instruments encode assumptions about risk, responsibility, and appropriate forms of evidence, thereby delimiting which aspects of a system are considered relevant for evaluation and which remain unarticulated~\cite{gebru2021datasheets, poirier2021reading, kitchin2016makes}. In public-sector contexts, structured disclosure mechanisms such as AI registers and impact assessments standardize how institutions report on algorithmic systems, often privileging completeness, consistency, and machine-readability over contextual detail or interpretive openness~\cite{kaushal2024automated, trujillo2025dsa}. Consequently, these mechanisms can create an appearance of accountability while leaving significant gaps in how decisions are understood and contested in practice~\cite{groesch2025big}.

These limitations are particularly consequential for individuals subject to algorithmic decision-making, who must interpret outcomes without access to the underlying decision logic. In such contexts, users often engage in collective sensemaking practices to reconstruct how systems operate~\cite{weick2005organizing, bucher2019algorithmic}. Studies of everyday interactions with algorithmic systems have shown how individuals develop ``folk theories" of algorithms--informal, experience-based models that people construct to explain how algorithmic systems function, make decisions, and respond to user behavior~\cite{rader2018explanations, ytre2021folk}. These folk theories are not necessarily accurate representations of system logic; rather, they are shaped by partial information, prior beliefs, and socially shared interpretations~\cite{eslami2016first, devito2017algorithms}. Through online forums and peer exchanges, individuals draw on others' experiences, compare outcomes, and iteratively refine these interpretations. Rather than relying on formal documentation, these practices produce situated, collectively negotiated understandings that approximate decision criteria through comparison, speculation, and inference.

Collective sensemaking thus emerges not as a peripheral phenomenon, but as a necessary response to the limits of institutional transparency~\cite{eiband2018bringing, rader2018explanations}. When accountability mechanisms fail to provide actionable interpretability, affected individuals turn to distributed forms of knowledge production to make sense of decisions, anticipate outcomes, and guide action~\cite{weick2005organizing, wenger1998communities}. In this process, algorithmic folk theories function as provisional epistemic frameworks that bridge the gap between institutional disclosures and lived experience, enabling individuals to act under conditions of uncertainty despite incomplete or opaque information~\cite{ananny2018seeing, devito2017algorithms, bucher2019algorithmic}. These practices are particularly salient in administrative contexts, where decisions are consequential yet opaque, requiring individuals to navigate systems whose evaluative criteria are only partially visible.

\subsection{Algorithmic Governance in Immigration and Visa Processing Systems}
Across jurisdictions, algorithmic decision-making has been increasingly adopted in immigration processes, including visa triage, risk assessment, fraud detection, and eligibility determination~\cite{molnar2018, zajko_automated_2023}. These systems operate in high-stakes contexts where decisions directly affect individuals' legal status, mobility, and access to rights and opportunities~\cite{sigona2012have}. However, scholars have raised concerns about their bias and discrimination, procedural fairness, and opacity, especially highlighting how applicants may struggle to understand, contest, or seek redress for decisions that shape their life trajectories~\cite{oluwasanmi2021algorithms}. The legal recourse for these applicants becomes more complicated due to the inherently transnational nature of immigration governance. Immigration systems evaluate individuals across national boundaries, incorporating factors such as nationality, travel history, and geopolitical context into decision-making processes~\cite{bigo2002security}. As a result, identical formal rules may produce uneven outcomes depending on applicants' positionality within global mobility regimes. Since these broader geopolitical and socio-economic structures are difficult to capture within formal governance frameworks~\cite{leurs2018five}, these dynamics complicate traditional notions of accountability, and algorithmic systems in such contexts can reinforce existing inequalities~\cite{belanger2025revisiting}.

In the Canadian immigration system, IRCC has deployed automated decision-support tools across multiple immigration streams, including systems for triaging TRV applications. It has used AIA as a mandatory ex-ante assessment tool~\cite{gc2026aiatool} and has published the AIA report, along with the National Research Council (NRC)'s peer review, through the Government of Canada's Open Government Portal~\cite{gc2025irccaia, NRC_visaalgo_review}. These are intended to function as disclosure mechanisms that make algorithmic systems more transparent and accountable to external audiences. Studies have shown that such impact assessments may rely on predefined categories of risk that do not fully capture the lived experiences of affected populations, and that their structured, often closed-ended formats can limit the extent to which contextual details about system operation and impact are documented~\cite{metcalf2021algorithmic, selbst2021institutional}. These critiques suggest that while AIAs play an important role in institutionalizing accountability practices, they may not fully reflect how algorithmic systems function in practice or how their impacts are experienced.

Emerging work in human-computer interaction (HCI) and computer-supported cooperative work (CSCW) further demonstrates how immigration systems are encountered through digital and algorithmic interfaces that mediate access to information, shape user experience, and reproduce structural inequalities. Studies of government-issued platforms, such as the US Citizenship and Immigration Services website, show how interface design, language choices, and navigation structures can reflect institutional priorities while creating barriers for diverse immigrant populations, including multilingual users and those with limited familiarity with bureaucratic systems~\cite{chen2025navigating}. These interfaces often serve as critical points of interaction where migrants must interpret complex legal and procedural information under uncertainty, with design choices shaping how rights, obligations, and possible actions are understood. Williams and Schoenebeck aptly conceptualized these systems as ``digital border walls"~\cite{williams2025african}, where interactions with online platforms become integral to the experience of migration governance, shaping not only access to services but also perceptions of inclusion, exclusion, and institutional legitimacy.

Algorithmic and data-driven infrastructures further extend these dynamics by structuring how applications are processed, documents are evaluated, and responses are generated. For instance, automation in immigration workflows, such as document classification and response-generation systems, illustrates how bureaucratic processes are increasingly reorganized around machine-readable categories and templated decision logic, with humans positioned in supervisory roles~\cite{mukherjee2020immigration}. At the same time, empirical studies of immigration surveillance technologies highlight how migrants experience these systems as opaque, intrusive, and difficult to interpret, often lacking clear information about how decisions are made or how technologies function~\cite{owens2025understanding}. Together, this body of work underscores that algorithmic migration governance is not only a matter of institutional design or policy compliance, but also a lived experience shaped by interfaces, infrastructures, and asymmetries in access to knowledge about how systems operate.

Overall, existing scholarship on migration governance has tended to examine governance frameworks, policy instruments, and system design in isolation, focusing on institutional design, interfaces, and technological infrastructures. However, there is a lack of research examining how applicants make sense of opaque decision-making processes, how their interpretations align or diverge from institutional representations, and how these dynamics unfold in transnational administrative contexts. This paper addresses this gap by analyzing IRCC's AIA and NRC's peer-review reports on the TRV triage system, alongside applicant discourse drawn from online communities.

%% file: sections/methods.tex
\section{Methods}
We adopt a comparative analytic design that treats (1) IRCC's AIA reports for its automated TRV triage system, and (2) a Reddit corpus of discussions on Canadian immigration processes as two sites through which accountability is articulated (as institutional artifacts) and interpreted (as applicant discourse). The AIA represents the government's formal articulation of risk, mitigation, and accountability regarding automated decision-making for TRV. Reddit discussions, in contrast, provide downstream accounts of how applicants interpret, reconstruct, and respond to institutional decisions amid uncertainty, delay, limited transparency, and perceived automation in IRCC processes. Pairing these sources enables us to examine how accountability mechanisms articulated in institutional documentation are translated--and where they break down--in applicants' experiences. We analyze these datasets through the lens of ADMAPS~\cite{saxena2021framework}, which conceptualizes public-sector algorithmic governance along three dimensions: bureaucratic processes, human discretion, and algorithmic decision-making. We use it to examine how these dimensions are formally articulated within IRCC's institutional documentation, i.e., the AIA report. We then analyze Reddit discussions to identify experiential dimensions that exceed or complicate this framework. This comparative approach enables us to identify experiential dimensions underrepresented in existing assessment instruments and to assess the adequacy of ADMAPS in transnational contexts.

\subsection{Data Collection and Overview}

\subsubsection{IRCC Algorithmic Impact Assessment}
We retrieved the AIA report titled \textit{Advanced Analytics Triage of Overseas Temporary Resident Visa Applications} from the Government of Canada's Open Government Portal\footnote{\url{https://open.canada.ca/data/en/dataset/6cba99b1-ea2c-4f8a-b954-3843ecd3a7f0}}. The report is structured as a standardized questionnaire used to evaluate risks associated with automated decision systems. It contains 65 questions on risks identified within the project, the system, the algorithm, the automated decision, the impact of the decision, and the data used, and 41 questions on steps taken to mitigate risk, including consultations with stakeholders and de-risking and mitigation measures. The assessment produces an overall risk score and a mitigation score. Although procedurally comprehensive, the document relies predominantly on closed-ended responses. Only nine of the 65 risk-related questions solicit a long-form explanation. As such, contextual details regarding system implementation, institutional constraints, or affected populations are limited within the report itself.

\subsubsection{Reddit Corpus}
Reddit is a large-scale online discussion platform organized into user-created topical communities (``subreddits"), where users create posts and threaded comments. To understand applicants' accounts, we constructed a corpus of Reddit discussions from Canadian visa and immigration-focused subreddits. We selected subreddits based on two criteria: (a) substantial membership size and sustained activity, and (b) primary topical focus on Canadian immigration or IRCC processes. Table~\ref{tab:subreddits} summarizes the included subreddits and the number of extracted posts. Smaller communities (e.g., r\textbackslash{CanadaVisa}) were included when highly aligned with overseas visa processes. We collected data on September 23, 2025, using the Python Reddit API Wrapper (PRAW). The earliest post collected in our dataset dates back to 13 June 2021. Deleted posts, removed comments, and content from banned users were not retrievable at the time of extraction and are therefore absent from the dataset. This limitation may disproportionately affect discussions involving sensitive experiences or contested decisions. We treat these discussions not as representative public opinion, but as situated accounts through which applicants collectively interpret institutional processes and outcomes.

\begin{table*}[!ht]
\centering
\caption{Reddit corpus: Immigration-focused subreddits}
\label{tab:subreddits}
\begin{tabular}{p{3.2cm}p{1.5cm}p{1.5cm}p{6cm}}
 \toprule
 Subreddit & Followers & Posts & Discussion Focus \\ 
 \midrule
 r\textbackslash{immigrationCanada} & 254k & 993 & Questions and discussion regarding immigration to Canada \\ 
 \hline
 r\textbackslash{CanadaImmigrant} & 10k & 952 & Experiences related to immigration \\ 
 \hline
 r\textbackslash{MovingToCanada} & 7.8k & 617 & Relocation assistance and advice \\ 
 \hline
 r\textbackslash{IRCCDiscussion} & 1.6k & 463 & IRCC processes and policy updates \\ 
 \hline
 r\textbackslash{CanadaVisa} & 1.1k & 202 & PR, work permit, study permit, and visitor visa discussions \\ 
 \bottomrule
\end{tabular}
\end{table*}

\subsection{Data Analysis}

\subsubsection{Analysis of the IRCC AIA and NRC Peer Review}
We conducted a deductive qualitative analysis of the IRCC AIA report and the NRC peer review of the TRV triage system. Given the predominance of binary responses, our analysis focused on how risk is framed, operationalized, and bounded within the assessment instrument. The first two authors conducted the initial open coding of the AIA report. Where applicable, we analyzed the binary responses jointly with their accompanying descriptive fields. This approach allows us to examine both the presence and absence of disclosure within the assessment document. We then mapped these open codes to the three dimensions of the ADMAPS framework~\cite{saxena2021framework}. The ADMAPS framework further conceptualizes algorithmic governance as emerging from the interaction of these elements: human discretion, bureaucratic processes, and algorithmic decision-making. Human discretion encompasses professional expertise, value judgments, and heuristic reasoning exercised by frontline practitioners; bureaucratic processes capture organizational conditions such as resources and constraints, administrative routines and training, and governing laws and policies; and algorithmic decision-making attends to the role of data, the form of decision support (e.g., predictive or prescriptive), and the degree of uncertainty embedded in these systems. The codes were revised, and their mappings were consolidated and finalized through discussions with the two supervising authors. This structured approach enables us to assess which dimensions are foregrounded and which are minimized in the assessment instrument. 


\subsubsection{Analysis of Reddit Discussions}
We used a mixed-methods approach in which computational clustering supported the identification of recurring discussion patterns, followed by in-depth qualitative analysis~\cite{corbin1990grounded} to interpret their meaning in relation to institutional processes. Such methodological combinations are widely adopted in social computing research to balance large-scale pattern detection with interpretive rigor~\cite{muller2016machine, baumer2017comparing}. 

First, we used BERTopic~\cite{grootendorst2022bertopic} to identify recurring themes across Reddit threads. We generated embeddings for the collected Reddit posts using the Qwen-3 language model~\cite{bai2023qwen}. The resulting 1024-dimensional embeddings were reduced using UMAP~\cite{mcinnes2018umap} to preserve local semantic structure prior to clustering. We selected HDBSCAN to generate clusters due to its robustness to variable cluster densities~\cite{mcinnes2017hdbscan}. We selected the hyperparameters through grid search, prioritizing topic diversity and minimizing the proportion of posts classified as noise. Automatically generated topic representations were treated as intermediate outputs rather than final analytic categories.

Next, the first two authors manually reviewed, merged, and interpreted clusters to ensure internal coherence and substantive interpretability. During this phase, we paid particular attention to identifying clusters that surfaced experiential dimensions not directly captured in the IRCC AIA reports or readily mappable onto the three ADMAPS dimensions. Then, we conducted an in-depth qualitative analysis of posts within those clusters. Following the grounded-theory approach, first, we generated open codes by identifying frequent topics, intents, and processes mentioned in the posts. Examples of open codes included ``guessing approval thresholds," ``comparing visa offices," and ``knowing about others' timelines." These open codes were subsequently grouped into axial codes by identifying conceptual relationships among them. For example, open codes such as ``guessing approval thresholds" and ``asking how much funds are enough" were merged into the axial category ``inferring hidden decision criteria." Finally, axial codes were consolidated into higher-level selective codes that captured broader structural patterns across discussions. For instance, axial codes related to inferring hidden criteria, nationality-based speculation, and collective information pooling were integrated into the selective code ``epistemic asymmetry," reflecting unequal access to institutional decision logic. The supervising authors were involved throughout the coding process, providing analytic feedback and helping resolve disagreements to ensure interpretive rigor. We conceptualized the unifying pattern in the selective codes as ``transnational asymmetry," which, when read alongside the institutional analysis, informed our identification of epistemic, jurisdictional, and temporal–relational asymmetries.



%% file: sections/results.tex
\section{Results}
We structure our results to examine how accountability in automated visa triage is articulated, translated, and experienced across institutional and applicant contexts. We first analyze IRCC's AIA report and NRC peer review to understand how automation is framed within bureaucratic and technical governance. We then examine Reddit discussions as downstream sites of applicant sensemaking, where individuals collectively interpret and respond to system outcomes. Reading these together reveals systematic asymmetries between institutional accountability mechanisms and their experiential uptake. Thus, rather than treating Reddit discussions as a standalone social computing phenomenon, we analyze them as downstream sites where applicants collectively interpret and compensate for institutional opacity.

\subsection{Institutional Artifacts Mapped to ADMAPS}\label{sec:aia}
We map two institutional artifacts: (1) IRCC's AIA for the TRV triage system and (2) the NRC peer review of that system, onto ADMAPS's three dimensions: bureaucratic processes, human discretion, and algorithmic decision-making~\cite{saxena2021framework}, to examine how institutional documentation articulates discretion, organizational practice, and model operation in relation to automated visa triage.

\subsubsection{Bureaucratic Processes}
The AIA operationalizes bureaucratic process through a compliance-oriented questionnaire and mitigation checklist. It assigned an impact level of 2 (Moderate impact) to IRCC's advanced analytics triage system for TRV applications. This level denotes systems with moderate, reversible impacts on individuals and requires mandatory measures, including peer reviews, consultation records, quarterly reviews, basic bias testing, human override, and recourse processes~\cite{gc2026aiatool}. The AIA report describes documented processes for bias testing, data quality resolution, Gender-based Analysis (GBA+) analysis~\cite{gc2025gba}, audit trails, change logs, access controls, and override logging in the IRCC TRV triaging system, while repeatedly noting that many of these materials are not publicly available. The AIA also mentions internal and external consultations (including academia, the Office of the Privacy Commissioner, and immigration lawyers).

The NRC report foregrounds bureaucratic processes through tooling, documentation, and reproducibility practices. It upholds other bureaucratic governance practices, such as routinized maintenance, performance monitoring, and procedural accountability. For example, it notes that modeling and deployment occur within the Statistical Package for Social Sciences (SPSS) software environment, which is described as stable but limited in flexibility. It repeatedly emphasizes documentation and reproducibility, describing the overall approach as ``excellent, simple and clear," and stating that reproducibility risks are minimized through software choice, documentation, and variable justification. It also endorses a three-month retraining cycle for the models and recommends monitoring retraining parameters. We discuss the procedural accountability of the IRCC TRV application triage system in the following subsections.

\subsubsection{Human Discretion}
The AIA formally retains human discretion in algorithmic decision-making for TRV applications. It states that the system automates only positive eligibility determinations and does not refuse applications. All refusals continue to be made by officers, and officers make the final decision on each application. Even in cases with automated positive eligibility determinations, applications are sent to an officer for admissibility screening, and officers may revisit eligibility if they encounter information that affects the determination. The AIA also states that officers will not be aware of the rules used for triage or automated determinations, nor will they receive information about the system's analysis, thereby insulating human discretion from direct algorithmic influence. To locate the procedural accountability, the NRC review depicts the routing of applications ``to the appropriate agents for processing" as a discretionary division of labor rather than wholesale replacement. The NRC peer review primarily evaluates modeling methodology rather than officer decision practices. However, it emphasizes organizational risks rooted in human factors (e.g., legal, public perception, security) alongside performance goals, highlighting that professional judgment and institutional priorities shape the deployment of automation.

\subsubsection{Algorithmic Decision-Making}
The AIA characterizes the system as sorting applications into tiers based on complexity and automating certain positive eligibility determinations. It states that the system uses personal information and draws from multiple data sources, and it reports that the algorithmic process is not difficult to interpret or explain. However, the AIA provides limited detail on what features are used, how complexity tiers are operationalized, or how routing decisions affect processing outcomes. We see a similar mismatch in impact assessment. While the triaging system was assigned an overall ``moderate" impact score, the AIA assesses its impacts on individuals' rights and freedoms as ``little to no impact" and justifies this by emphasizing triage use, positive-only automation, and alignment with legislative/regulatory requirements. 

The NRC review complements the AIA report by providing concrete technical details about the model. It identifies the modeling technique as a decision tree and explicitly praises it for transparency. It also explains the training/testing design, class imbalance (refusals as a relatively small class), under-sampling approaches, and validation sampling. The report recommended experimentation with more complex model types (e.g., Random Forest, support vector machine, neural networks). However, IRCC responses further indicate that decision tree-based models were retained due to explainability trade-offs. This positions interpretability and procedural accountability as primary design constraints that shape algorithmic decision-making.

\subsection{Reddit Discussions as Structured Applicant Discourse}\label{sec:reddit}
\begin{figure}[!ht]
    \centering
    \includegraphics[width=0.95\linewidth]{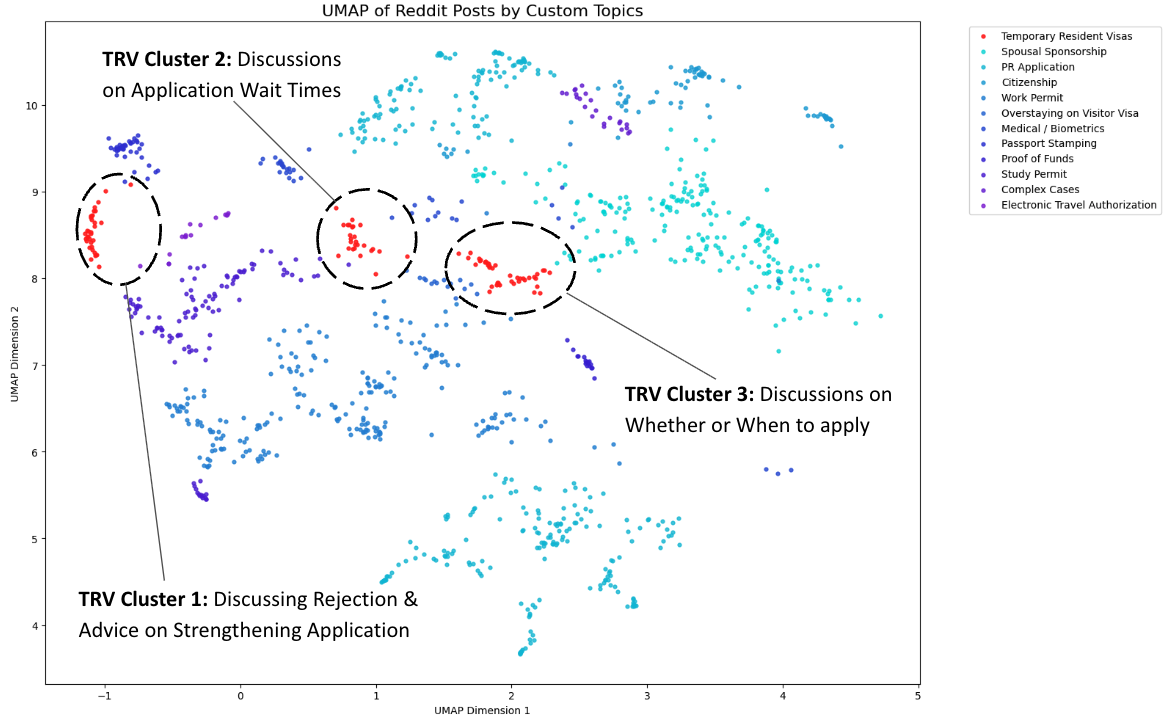}
    \caption{UMAP projection of Reddit posts using BERTopic embeddings. Three TRV-related clusters are highlighted: Cluster 1 (refusals and strengthening applications), Cluster 2 (processing delays and wait times), and Cluster 3 (eligibility, timing, and documentation strategy).}
    \label{fig:topic_model}
\end{figure}

Whereas institutional artifacts like the AIA and the NRC review describe how accountability in the automated TRV triage system is designed, Reddit discussions reveal how accountability is interpreted, reconstructed, and contested in practice. Using BERTopic, we identified 42 clusters after excluding outliers. Figure~\ref{fig:topic_model} shows the UMAP projection of posts. Manual inspection of these clusters revealed three of these being directly related to TRV processes: (1) refusal reasons and application strengthening, (2) processing times and delays, and (3) eligibility timing and document strategy. They represented about 5\% of the total corpus. The remaining clusters were centered on other immigration streams (e.g., spousal sponsorship, permanent residence, citizenship applications) or on procedural steps (e.g., biometrics) that cut across multiple visa categories. Because our institutional analysis focuses on the TRV triage system, we limited subsequent qualitative analysis to clusters uniquely associated with TRV applications. Clusters addressing broader immigration categories or cross-cutting procedural requirements were therefore excluded to maintain analytic alignment between institutional artifacts and applicant discourse. Rather than reflecting random discussion, the clusters related to TRV correspond to recurring stages of applicant interaction with the visa system and also organize around recurring uncertainties: how to present an application, how long processing will take, and how eligibility rules will be interpreted in specific circumstances.

\subsubsection{Cluster 1: Refusal Reasons and Application Strengthening}
Cluster 1 centers on speculating reasons for refusal and strategies for strengthening applications. Posts frequently include detailed financial figures, employment documentation, property ownership descriptions, and family ties. These discussions are structured to evaluate sufficiency with respect to funds, family and social ties, and documentation. Applicants describe rejection letters citing generic grounds such as insufficient ties or inadequate financial resources, and then attempt to determine which specific elements were deficient. For example, a user quoted texts from the received rejection letter their friend received:

\begin{quote}
    One friend had to apply for a transit visa for Canada. On the reasons why it says: ``I am not satisfied that you will leave Canada at the end of your stay as required by paragraph 179(b) of the IRPR\footnote{IRPR = Immigration and Refugee Protection Regulations}. ... I am refusing your application because you have not established that you will leave Canada, based on the following factors: your assets and financial situation are insufficient to support the stated purpose of travel." We are only passing through [two Canadian cities] to [a Japanese city], with about 4 hours of layover. He reported having C\$1500 for the stay in Canada. Is this not enough? Obviously, he has more money for the entire trip, but the application asked about the transit.
\end{quote}

Rather than outright disputing the legitimacy of refusal, posts often seek interpretive clarification. Users respond with heuristic advice: demonstrate property ownership, provide employment leave letters, avoid signaling future immigration intent, and ensure documentation aligns with the declared duration of stay. The conversational pattern resembles peer-led case review, where community members approximate institutional evaluation criteria through comparative anecdotes.

\subsubsection{Cluster 2: Processing Delays and Temporal Monitoring}
Cluster 2 is structured around uncertainty in waiting and processing. Posts frequently reference official IRCC processing times and contrast them with personal timelines. Let's consider the following conversation snippet:

\begin{quote}
    P1: The processing time has increased from around 210 days to over 390 days. So, I don't think I'll be hearing back from them anytime soon.

    P2: The processing time is the average of all applications processed, so I assume it should be half the wait time for normal visa applications.

    P3: The processing time has increased to 400 days, so I don't expect to hear back for another 3 months.

    P4: I just received it last week. It took 270 days to obtain it.
\end{quote}

These exchanges reveal a pattern of collective temporal monitoring. Here, the official IRCC estimate of processing times serves as a benchmark rather than a reliable predictor. Users refer to the publicly posted processing time and translate it into individualized forecasts. A key rationale to reinterpret the institutional metric is that it represents an average across all cases and may not apply uniformly. In this process, actual timelines serve as peer-based counterexamples that recalibrate others' expectations. Thus, rather than treating the official estimate as determinative, participants collectively triangulate expected outcomes through comparison, reinterpretation, and peer testimony. As institutional averages are re-contextualized through experiential data shared within the community, processing delays are simultaneously reframed as more than mere administrative inconvenience. Reddit posts regarding TRV applications often reference weddings, parental visits, academic programs, employment transitions, and travel plans. Consider the following conversation snippets from various threads:

\begin{quote}
    P5: I'm planning to apply for a second permit from outside Canada since I received a time-sensitive job offer.

    P6: I applied for my parents' and brother's visitor visas back in January, hoping they could come stay with me for 2-3 months over the summer. Unfortunately, their applications were refused. ... A few months later, I reapplied this time just for my parents since I found out I was pregnant, and the reason for their visit was no longer tourism, but their presence and support. ... This happened around the time of the political tensions, when IRCC was apparently rejecting most visitor visas unless it was for humanitarian reasons. ... My due date is in November, so I really need their support, but I'm worried another refusal could affect future chances. My question is: Should I apply for their super visa\footnote{Super visa is a multi-entry TRV option available for parents and grandparents of Canadian citizens or permanent residents.} now, or wait until next year?
\end{quote}

Applicants’ accounts show that TRV decisions are experienced not as discrete administrative outcomes but as temporally unfolding constraints that shape employment timelines and intimate family arrangements. As individuals navigate time-sensitive job offers or seek parental support during pregnancy, waiting and uncertainty become relationally embedded, structuring decisions across work, care, and mobility. Hence, waiting becomes a structuring condition that shapes life decisions beyond the application itself. For those affected by an unfavorable decision, the ``little to no" or ``moderate" impact assessment of IRCC's TRV triage in the institutional accountability artifacts does not feel relatable.

\subsubsection{Cluster 3: Eligibility, Timing, and Application Strategy}
Cluster 3 focuses on determining whether and when to apply, especially in complex status situations. Posts frequently involve overlapping permit types (e.g., study permit, post-graduation work permit (PGWP), visitor visa) and concerns about document validity. For example, users often asked whether holding a visa from another country might influence processing speed:

\begin{quote}
    What nationality are you? Do you think having a US visa will expedite things?
\end{quote}

Here, the application strategy itself is treated as a decision variable. Applicants deliberate over family composition, passport type, visa history, and intended travel duration. The system is experienced not only as a set of eligibility rules but as an evaluative apparatus sensitive to contextual signals. A user describes their personal situation as follows:

\begin{quote}
    I am an Afghan living in Kuwait. I previously traveled to a couple of countries for tourism. My recent trips were to the UK and Turkey. I am considering applying for a TRV to Canada. ... I am married and have kids. Should we all apply together, or should I apply on my own? How much bank balance should I show?
\end{quote}

Across clusters, applicants rarely describe directly interacting with the inner algorithms or components of algorithmic systems, which were prioritized in the AIA report and NRC review. Instead, they describe interacting with institutional outcomes, refusals, delays, and routing decisions, whose internal logic remains partially opaque. While the AIA and NRC reviews emphasize interpretability and procedural accountability in system design and organizational oversight, these qualities do not translate into interpretability or transparency for the applicant experience. For applicants, the system does not appear as a documented decision tree with audit trails and retraining cycles, but as a sequence of outcomes whose evaluative criteria and temporal dynamics remain unclear.

\subsection{Emergent Asymmetries Between Institutional Design and Applicant Experience}
Juxtaposing institutional artifacts with applicant discourse reveals not simply differences in perspective, but systematic failures in how accountability mechanisms translate across contexts. The AIA and the NRC review frame automation in TRV application processing as interpretable, procedurally governed, and bounded in impact through retained discretion in algorithm-supported and human-in-the-loop decision making, documented procedural safeguards, the interpretability of modeling techniques, and bounded-impact classifications. In contrast, applicants encounter a system whose evaluative logic, spatial positioning, and temporal consequences remain uncertain and uneven, especially around evaluative criteria, spatial positioning, and temporal exposure. These differences do not indicate that the system lacks documentation or oversight. Rather, they reveal asymmetries in the prioritization of interpretability aspects and impact factors across institutional and applicant positions. We identify three interrelated asymmetries: epistemic, jurisdictional, and temporal–relational.

\subsubsection{Epistemic Asymmetry: Unequal Access to Decision Logic}
Under ADMAPS, algorithmic decision-making is evaluated with respect to model choice, explainability, validation, and reproducibility. While the NRC review and AIA report explicitly praise the TRV triage system for its transparency, interpretability, audit trails, and documentation, interpretability does not manifest as an accessible understanding of evaluative criteria in applicant discourse. Instead, applicants attempt to reconstruct thresholds retrospectively. In Cluster 1, users debate whether a certain amount of bank balance is ``enough," whether property ownership sufficiently demonstrates ties to their home countries, and how documentation aligns with the IRPR guidelines. While the AIA explicitly references bias testing in relation to gender (e.g., through GBA+), it does not substantially address how other demographic and socioeconomic attributes, such as nationality and financial solvency, may structure decision-making. In contrast, discussions in Cluster 3 frequently center on whether nationality, prior visa history, or family composition influence approval outcomes. This divergence indicates that applicants perceive demographic positioning as central to fairness and accountability in TRV applications, even where such dimensions are not foregrounded in institutional documentation.

The institutionally validated transparency indicators and initiatives, such as model choice, documentation, and retraining, do not eliminate uncertainty at the point of application. In the absence of access to the system's decision logic, applicants lack the interpretive resources needed to make sense of outcomes--a form of hermeneutical disadvantage. As a result, they approximate evaluative criteria through peer testimony and anecdotal comparison, collectively reconstructing a logic that remains opaque within formal documentation. We describe this gap as an \emph{epistemic asymmetry}: interpretability is secured at the level of institutional governance, and applicants construct parallel interpretive practices through peer exchange, yet the system's authoritative evaluative logic remains structurally out of reach for those who are subject to its decisions. This indicates that existing transparency and explainability mechanisms operate primarily at the level of institutional auditability, rather than supporting meaningful interpretability for affected individuals.

\subsubsection{Jurisdictional Asymmetry: Spatial Positioning and Sovereign Exposure}
The AIA and NRC documents treat the system as an institutional mechanism applied consistently within regulatory boundaries. However, applicant discussions reveal sensitivity to geographic and nationality-based positioning. For example, in Cluster 3, users ask whether holding a US visa might ``expedite things," whether passport type or applying from certain visa centers influences processing, and whether applying together or separately alters evaluation. These deliberations reflect an awareness that identical formal rules may produce different outcomes depending on national origin, visa history, or country of application, due to factors such as the volume of TRV applications in that category.

Institutionally, algorithmic triage is framed as routing applications by complexity. Experientially, applicants interpret routing and waiting as shaped by geopolitical positioning. While documentation emphasizes uniformity and procedural alignment with legislative criteria, applicants perceive exposure to the system as uneven across national contexts. We describe this as \emph{jurisdictional asymmetry}: applicants encounter the system from varied jurisdictional positions that shape their vulnerability to delay, scrutiny, or routing—differences not foregrounded in institutional documentation. As a result, accountability mechanisms grounded in uniform regulatory assumptions fail to account for uneven exposure across transnational populations.

\subsubsection{Temporal–Relational Asymmetry: Waiting as Distributed Consequence}
The TRV triage system is characterized as efficiency-enhancing and assistive, with positive eligibility automation intended to streamline processing. Although the AIA categorizes impacts on rights and freedoms as ``little to no impact" and frames refusals as reversible, Reddit discussions demonstrate that waiting time functions as a central experiential consequence, revealing a divergence between institutional impact classification and lived temporal exposure. For example, as posts in Cluster 2 reference weddings, parental visits, care needs during pregnancy, academic enrollment, employment transitions, and family reunification, we found that Reddit users often discussed how the temporal uncertainty of the TRV application processing extends beyond the application itself, shaping relational and life-planning decisions. We describe this as \emph{temporal–relational asymmetry}: although institutional impact assessments classify effects as moderate and reversible, applicants experience prolonged temporal exposure as consequential, deeply entangled with relational commitments, and unevenly borne across individuals. This suggests that current impact assessments systematically under-theorize time as a site of harm, particularly in contexts where administrative delays are socially and relationally consequential.

Taken together, these asymmetries reveal a limitation in the ADMAPS framework: while it captures accountability within institutional boundaries, it does not account for how those mechanisms are translated, experienced, and redistributed across transnational contexts. We therefore extend ADMAPS to incorporate this dimension, foregrounding how accountability is not only designed, but differentially encountered by the populations subject to it.

%% file: sections/discussion.tex
\section{Discussion}
In this section, we first show how accountability is collectively reconstructed under conditions of opacity, then examine how administrative assessments overlook temporal--relational burdens, and finally extend these insights to transnational contexts, where accountability is mediated by applicants' uneven positioning across borders.

\subsection{Accountability as Collective Reconstruction Under Opacity}
Our analysis highlights structural limitations of the AIA as a transparency mechanism. The predominance of binary (yes/no) questions constrains the extent to which institutions can provide detailed, contextualized explanations of system behavior. As a result, many claims regarding fairness, bias mitigation, and procedural safeguards remain internally asserted but externally unverifiable. We found how accountability can be actively reconstructed through collective sensemaking practices among applicants when it is limited by institutional mechanisms.

While accountability recommendations often emphasize documentation and auditability as key mechanisms for enabling oversight and fairness~\cite{binns2018algorithmic, diakopoulos2016accountability, mitchell2019model}, applicants interacting with the TRV system do not encounter algorithmic decision-making as documented models or audit trails, but as opaque outcomes--refusals, delays, and routing decisions--whose underlying logic, often relying on institutional jargon, remains inaccessible. Thus, the AIA report and the NRC peer review about the IRCC's TRV triage system function more as institutional reporting than as mechanisms for enabling meaningful public understanding. In response, they engage in collective sensemaking practices, using platforms such as Reddit to reconstruct evaluative criteria through peer comparison, anecdotal evidence, and heuristic reasoning. These discussions frequently center on common questions (e.g., evidence of financial adequacy) in the TRV application and speculate how the responses to those questions and their representation influence outcomes.

These practices reflect attempts to approximate decision logic, develop situated interpretations and explanations, and lead to various folk theories that guide future applicants in preparing their TRV applications. As institutional artifacts aimed at accountability and transparency lack interpretability for general applicants, this increases their reliance on collective sensemaking through online platforms. Applicants with more common or typical cases benefit from shared community knowledge, whereas those with atypical or complex circumstances are less likely to receive useful guidance, resulting in unequal knowledge sharing within the applicant population. Moreover, while these discussions can inform applicants’ preparation strategies, they also introduce new dynamics into the system. In the absence of interpretable transparency and accountability measures, attempts to align with perceived decision criteria may encourage strategic presentation or misrepresentation of information, increasing the verification burden on IRCC adjudicators and, in turn, contributing to additional scrutiny and delays for future applicants.

\subsection{Administrative Assessment based on Temporal--Relational Exposure}
Public-sector algorithmic systems in administrative infrastructures are often justified through efficiency, standardization, and procedural safeguards~\cite{saxena2021framework, redden2022automating}. Within this framing, systems such as IRCC's automated triage are evaluated in terms of their ability to improve processing efficiency, manage application volumes, and preserve human discretion. While it is characterized as a system with a moderate impact based on the internal assessment score, our findings complicate this view by showing that these systems and their impacts are experienced not merely as administrative tools, but as temporally and relationally distributed processes whose consequences extend beyond institutional boundaries. Particularly, processing time emerges not merely as a technical metric but as a structuring condition shaping applicants' life decisions, events, and opportunities. While the AIA characterizes the triage system as having ``moderate" or even minimal impact on individuals' rights and freedoms, applicants experience prolonged waiting and uncertainty as consequential and deeply embedded in their temporal-relational contexts. This reveals how institutional impact classifications fail to account for how time itself functions as a distributed burden.

Furthermore, the efficiency gains associated with automated triage are not evenly distributed. Because the system prioritizes ``routine" applications for streamlined processing and automated positive determinations, those already positioned as low-risk benefit most from reduced wait times. In contrast, applications requiring additional verification or deemed more complex continue to experience longer delays. As a result, efficiency operates asymmetrically: it accelerates already straightforward cases while offering limited relief for those facing the greatest uncertainty and stakes. This finding extends prior work by showing that administrative efficiency, when mediated through algorithmic triage, can reproduce uneven experiential outcomes. The institutional impact assessment artifacts also do not outline what steps need to be taken to uniformly improve service.

\subsection{Transnational Asymmetry in Accountability Across Borders}
Algorithmic systems in migration and border governance have been widely critiqued for their implications on mobility, surveillance, and differential access to rights~\cite{amoore2006biometric, leurs2018five, bigo2002security}. Applicants frequently speculate about whether uneven geopolitical factors, such as nationality, country of application, or prior travel history, influence processing outcomes. These concerns point to a broader jurisdictional asymmetry, in which individuals encounter the same formal system from uneven positions within global mobility regimes. Although institutional documentation emphasizes procedural consistency and standardized criteria~\cite{selbst2021institutional}, applicants perceive their exposure to the system as differentiated--shaped by application volumes, geopolitical relationships, and the perceived risk associated with specific regions. This asymmetry is further reflected in disparities in processing times across countries. While such differences may stem from variations in application complexity, verification procedures, or policy mandates, they are experienced by applicants as uneven access to mobility and opportunity. Importantly, these experiential disparities remain largely backgrounded in institutional accounts, which foreground standardization and alignment with regulatory criteria.

Taken together, these findings suggest that existing frameworks for evaluating algorithmic governance--such as ADMAPS--are limited in their ability to account for how systems operate across jurisdictional boundaries. While they effectively capture institutional dimensions such as bureaucratic processes, human discretion, and model design, they do not fully address how these systems are experienced by individuals situated outside the governing state. This limitation points to the need for conceptual frameworks that incorporate transnational dimensions of algorithmic governance, particularly in domains such as immigration, where decisions are inherently cross-border and unevenly distributed.

While the ADMAPS framework effectively captures institutional dimensions of automated governance within organizational boundaries~\cite{murad2021beyond, haataja2020public, saxena2021framework}, our study demonstrates that when extended to transnational domains such as visa adjudication, these dimensions do not fully account for how systems are experienced by individuals positioned outside the governing state. In such contexts, institutional safeguards within a country do not translate uniformly into meaningful accountability for affected populations in others. Instead, they generate uneven experiential effects across borders, revealing a gap between institutional articulation and lived experience.

To address this limitation, we propose an extension of ADMAPS by introducing a fourth dimension--\emph{transnational asymmetry}--thereby transforming it into Algorithmic Decision-Making for Administrative Transnational Systems (ADMATS) (see Figure~\ref{fig:modified_admaps}). This added dimension captures how accountability is differentially produced and experienced across borders, foregrounding the structural conditions under which individuals encounter algorithmic decision-making systems. Empirically, transnational asymmetry comprises three interrelated components:

\begin{figure}[!ht]
    \centering
    \includegraphics[width=\linewidth]{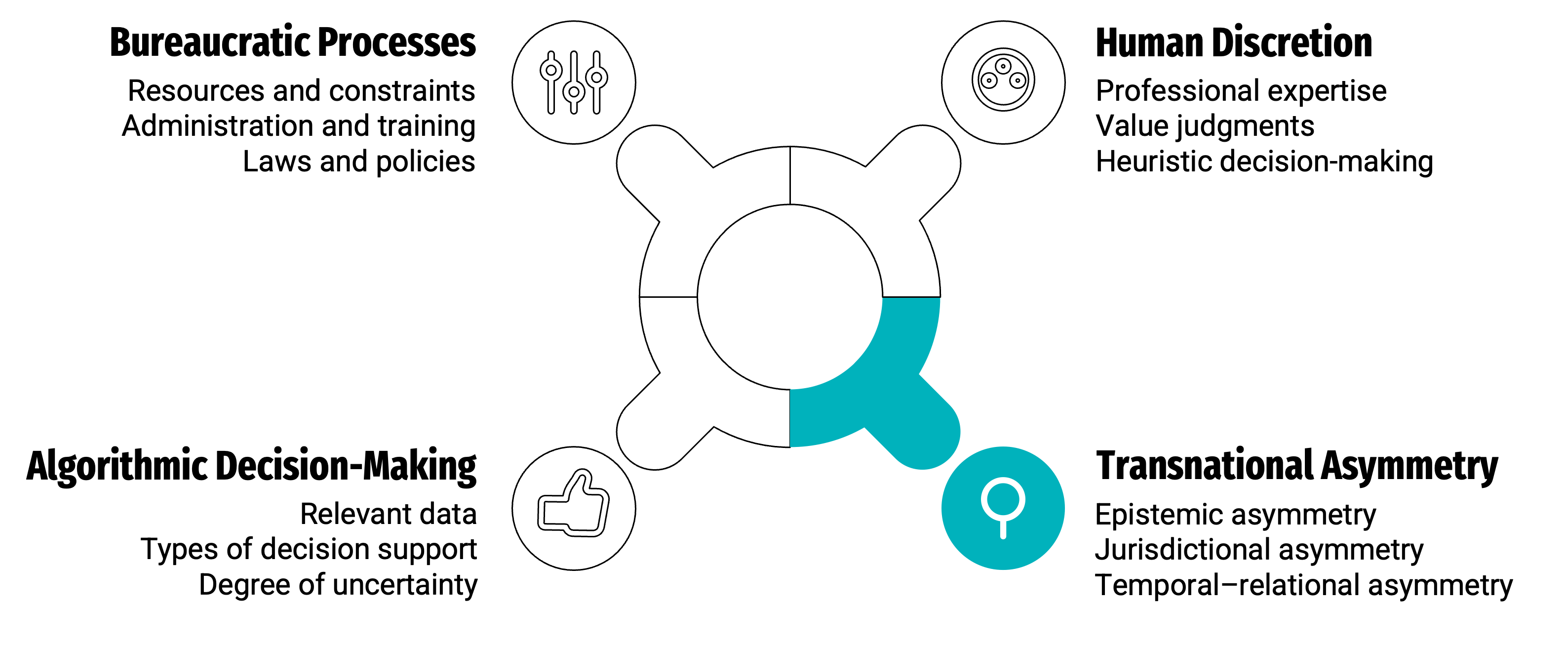}
    \caption{A Framework for Algorithmic Decision-Making in Administrative Transnational System (ADMATS). The dimensions of the Algorithmic Decision-Making Adapted for the Public Sector (ADMAPS) framework are shown in black and white. The additional dimension introduced in ADMATS is highlighted in turquoise.}
    \label{fig:modified_admaps}
\end{figure}

\begin{itemize}[leftmargin=*]
    \item \textbf{Epistemic asymmetry}: Institutional interpretability--such as documented decision rules, audit trails, and validation procedures--does not equate to accessible procedural transparency or accountability for applicants. Despite being familiar with different terminologies, norms, and contexts, they often lack insight into evaluative criteria described and explained differently.
    
    \item \textbf{Jurisdictional asymmetry}: Although algorithmic triage is framed as uniformly applied, applicants experience differentiated exposure shaped by their geopolitical positioning, regional classification, and perceived risk profiles.
    
    \item \textbf{Temporal--relational asymmetry}: Institutional impact categorizations do not capture how applicants experience the consequences of decisions in relation to their temporal constraints (e.g., urgency, delays) and relational dependencies (e.g., family, employment, migration trajectories).
\end{itemize}

ADMATS does not replace ADMAPS; rather, it extends it by incorporating transnational dimensions into the evaluation of administrative algorithmic systems operating across borders. Under ADMAPS alone, systems such as TRV triage may appear procedurally accountable, discretion-preserving, and technically interpretable. ADMATS demonstrates that these properties can coexist with asymmetries in interpretive access, jurisdictional vulnerability, and temporal burden. By incorporating transnational asymmetry as a fourth dimension, ADMATS enables a more comprehensive analysis of how administrative algorithmic systems interface with conditions under which those decisions are understood, contested, and lived by individuals situated across uneven global contexts. 

This is intended for administrative algorithmic systems whose decisions or triage functions cross jurisdictional boundaries and whose accountability mechanisms are institutionally produced in one jurisdiction but experienced by affected populations across others. It is less applicable to transnational systems that are primarily commercial, recommender-based, or infrastructural, without an administrative adjudicatory function, unless those systems become embedded in public-sector eligibility, enforcement, mobility, or resource allocation decisions. Thus, ADMATS is not a general theory of all transnational algorithms, but rather a framework for analyzing cross-border administrative algorithmic governance in which institutional accountability, affected-person interpretability, jurisdictional positioning, and temporal consequences diverge.

%% file: sections/conclusion.tex
\section{Conclusion}
Our findings show that existing governance instruments for public-sector AI, such as the AIA, fall short not because of their absence, but because accountability is unevenly interrogated and understood across bureaucratic and experiential contexts. Addressing this requires moving beyond compliance-oriented reporting toward mechanisms that are interpretable and actionable for those affected. Governance instruments and impact evaluations must incorporate the lived experiences of applicants, particularly those in intersectional institutional jurisdictions, to surface epistemic and structural asymmetries that current processes overlook. Public institutions' claims about fairness and impact must be both accurate and perceived as meaningful by the publics they govern. Taken together, these shifts reframe algorithmic accountability as not only a matter of institutional structure and procedure, but of how accountability is contextually experienced and distributed.